\documentclass[a4paper,11pt]{article}
\usepackage{pos}
\usepackage{braket}
\usepackage{sidecap}
\sidecaptionvpos{figure}{m}
\sidecaptionvpos{table}{m}

\usepackage{adjustbox}
\usepackage{cleveref}

\usepackage{natbib}
\newcommand{\Esym}{\braket{E_{\mathrm{sym}}(t)}}
\newcommand{\wsym}{w_0^{\mathrm{sym}}}

\newcommand{\tsym}{t_0^{\mathrm{sym}}}
\newcommand{\ssym}{s_0^{\mathrm{sym}}}

\title{Gradient flow scale-setting with $N_f=2+1+1$ Wilson-clover twisted-mass fermions}

\author[a,b]{C.~Alexandrou}
\author[b]{S.~Bacchio}
\author[c]{G.~Bergner}
\author[d]{P.~Dimopoulos}
\author[b]{J.~Finkenrath}
\author[e]{R.~Frezzotti}
\author[f,g]{M.~Garofalo}
\author*[h]{B.~Kostrzewa}
\author[b]{G.~Koutsou}
\author[i]{P.~Labus}
\author[j]{F.~Sanfilippo}
\author[j]{S.~Simula}
\author[f]{M.~Ueding}
\author[f]{C.~Urbach}
\author[k]{U.~Wenger}

\emailAdd{kostrzewab@informatik.uni-bonn.de}

\affiliation[a]{Department of Physics, University of Cyprus, 20537~Nicosia, Cyprus}
\affiliation[b]{Computation-based Science and Technology Research Center, The Cyprus Institute, 20~Konstantinou Kavafi Street, 2121~Nicosia, Cyprus}
\affiliation[c]{University of Jena, Institute for Theoretical Physics, Max-Wien-Platz 1, D-07743~Jena, Germany}
\affiliation[d]{Dipartimento di Scienze Matematiche, Fisiche e Informatiche, Universit{\`a} di Parma  and  INFN, Gruppo Collegato  di  Parma, Parco  Area  delle Scienze  7/a (Campus), I-43124~Parma, Italy}
\affiliation[e]{Dipartimento di Fisica, Universit{\`a} di Roma ``Tor Vergata" and INFN, Sezione di Tor Vergata, Via della Ricerca Scientifica 1, I-00133~Roma, Italy}
\affiliation[f]{HISKP (Theory), Rheinische Friedrich-Wilhelms-Universit{\"a}t Bonn, Nussallee~14-16, D-53115~Bonn, Germany}
\affiliation[g]{Dipartimento di Fisica, Universit{\`a} Roma Tre and INFN, Sezione di Roma Tre, Via della Vasca Navale 84, I-00146 Rome, Italy}
\affiliation[h]{High Performance Computing and Analytics Lab, Rheinische Friedrich-Wilhelms-Universit{\"a}t Bonn, Friedrich-Hirzebruch-Allee~8, D-53115 Bonn, Germany}
\affiliation[i]{Fraunhofer Institute for Industrial Mathematics (ITWM), Fraunhofer-Platz~1, D-67663~Kaiserslautern, Germany}
\affiliation[j]{Istituto Nazionale di Fisica Nucleare, Sezione di Roma Tre, Via della Vasca Navale 84, I-00146~Rome, Italy}
\affiliation[k]{Institute for Theoretical Physics, Albert Einstein Center for Fundamental Physics, University of Bern, Sidlerstrasse~5, CH-3012 Bern, Switzerland}

\abstract{We present a determination of the gradient flow scales $w_0$, $\sqrt{t_0}$ and $t_0/w_0$ in isosymmetric QCD, making use of the gauge ensembles produced by the Extended Twisted Mass Collaboration (ETMC) with $N_f=2+1+1$ flavours of Wilson-clover twisted-mass quarks including configurations close to the physical point for all dynamical flavours. The simulations are carried out at three values of the lattice spacing and the scale is set through the PDG value of the pion decay constant, yielding $w_0=0.17383(63)$ fm, $\sqrt{t_0}=0.14436(61)$ fm and $t_0/w_0=0.11969(62)$ fm. Finally, fixing the kaon mass to its isosymmetric value, we determine the ratio of the kaon and pion leptonic decay constants to be $f_K/f_{\pi}=1.1995(44)$.}

\FullConference{The 38th International Symposium on Lattice Field Theory, LATTICE2021
  26th-30th July, 2021
  Zoom/Gather@Massachusetts Institute of Technology
}

\begin{document}
\maketitle

\section{Introduction}

Precise and accurate scale setting is of central importance as lattice QCD calculations target high precision determinations of the hadron spectrum, the nucleon axial radius, precision inputs for electroweak tests of the Standard Model or the hadronic contribution to the muon $g-2$.
The lattice scale may enter either relatively to compare calculations at different values of the inverse bare coupling $\beta = 6/g_0^2$, indirectly when used to fix other bare parameters of the theory such as the quark masses or as an absolute scale in the conversion of dimensionful observables to physical units.
Depending on the case, its uncertainty either indirecty or directly also propagates to the error estimates of the final results of a given calculation.

The gluonic scales $t_0$~\cite{Luscher:2010iy} and $w_0$~\cite{Borsanyi:2012zs} have been employed widely as intermediate scales~\cite{Bruno:2013gha,Dowdall:2013rya,Bornyakov:2015eaa,HotQCD:2014kol,RBC:2014ntl,MILC:2015tqx,Bruno:2016plf,Miller:2020evg,Hollwieser:2020qri,Borsanyi:2020mff} and have also previously been studied specifically in the context of the ETMC~\cite{Deuzeman:2012jw,ETM:2015ned,Alexandrou:2018egz}.
They are attractive because they are comparatively easy to calculate with high statistical precision, do not involve complicated fitting procedures and can easily be integrated into the ensemble production workflow.
In this contribution we give some details on our current determinations of these scales and also present our recent calculation of $f_K/f_\pi$~\cite{ExtendedTwistedMass:2021qui}. 
We also make use of the scales in the calculation of quark masses from mesonic inputs~\cite{ExtendedTwistedMass:2021gbo,Alexandrou:2021wwd} as well as leptonic meson decay constants~\cite{Dimopoulos:2021qsf}.

\begin{table}
\centering
\begin{adjustbox}{width=0.7\textwidth}
\begin{tabular}{||c|c|c|c|c||c|c|c||}
\hline
 ensemble & $\beta$ & $V / a^4$ & $a~\mbox{(fm)}$ & $a \mu_\ell$ & $M_\pi~\mbox{(MeV)}$ & $L~\mbox{(fm)}$ & $M_\pi L$\\
\hline
cA211.53.24& $1.726$ & $24^3 \times~48$ & $~0.0947~(4)~$ & $~0.00530~$ & $~346.4~(1.6)~$ & $2.27$ & $3.99$\\
cA211.40.24&               & $24^3 \times~48$ &                              & $~0.00400~$ & $~301.6~(2.1)~$ & $2.27$ & $3.47$\\
cA211.30.32&               & $32^3 \times~64$ &                              & $~0.00300~$ & $~261.1~(1.1)~$ & $3.03$ & $4.01$\\
cA211.12.48&               & $48^3 \times~96$ &                              & $~0.00120~$ & $~167.1~(0.8)~$ & $4.55$ & $3.85$\\
\hline
cB211.25.24& $1.778$ & $24^3 \times~48$ & $~0.0816~(3)~$ & $~0.00250~$ & $~259.2~(3.0)~$  & $1.96$ & $2.57$ \\
cB211.25.32&               & $32^3 \times~64$ &                              & $~0.00250~$ & $~253.3~(1.4)~$  & $2.61$ & $3.35$\\
cB211.25.48&               & $48^3 \times~96$ &                              & $~0.00250~$ & $~253.0~(1.0)~$  & $3.92$ & $5.02$\\
cB211.14.64&               & $64^3 \times~128$ &                              & $~0.00140~$ & $~189.8~(0.7)~$  & $5.22$ & $5.02$\\
cB211.072.64&               & $64^3 \times~128$ &                              & $~0.00072~$ & $~136.8~(0.6)~$  & $5.22$ & $3.62$\\
\hline
cC211.06.80& $1.836$ & $80^3 \times 160$ & $~0.0694~(3)~$ & $~0.00060~$ & $~134.2~(0.5)~$ & $5.55$ & $3.78$\\
\hline
\end{tabular}
\end{adjustbox}
\caption{\it \footnotesize Overview of the light quark bare mass, $a \mu_\ell = a \mu_u = a \mu_d$, of the pion mass $M_\pi$, of the lattice size $L$ and of the product $M_\pi L$ for the various ETMC gauge ensembles used in this work. The values of the lattice spacing $a$ and the values of $M_\pi$ and $L$ correspond to the absolute scale $w_0 = 0.17383(63)$ fm. \label{tab:ensembles}}
\end{table} 

\section{Lattice Setup and Statistical Properties}

We make use of $N_f=2+1+1$ flavours of Wilson-clover twisted-mass fermions tuned to maximal twist, ensuring automatic $\mathcal{O}(a)$-improvement of all physical observables~\cite{Frezzotti:2001ea,Frezzotti:2003ni}.
We employ the tmLQCD software suite~\cite{Jansen:2009xp,Abdel-Rehim:2013wba,Deuzeman:2013xaa} linked against an extended version of the QPhiX~\cite{QPhiX-github,10.1007/978-3-642-38750-0_4,Schrock:2015gik,joo2016optimizing,joo2015wilson,7012993} library as well as DD$\alpha$AMG~\cite{Frommer:2013fsa,Frommer:2013kla,Alexandrou:2016izb,Alexandrou:2018wiv}.
Details of our ensembles are given in \Cref{tab:ensembles} and we refer to Refs.\cite{Alexandrou:2018egz,ExtendedTwistedMass:2021qui,Finkenrath:2021wwd} for details on their generation and the corresponding algorithmic setup.

\begin{SCfigure}
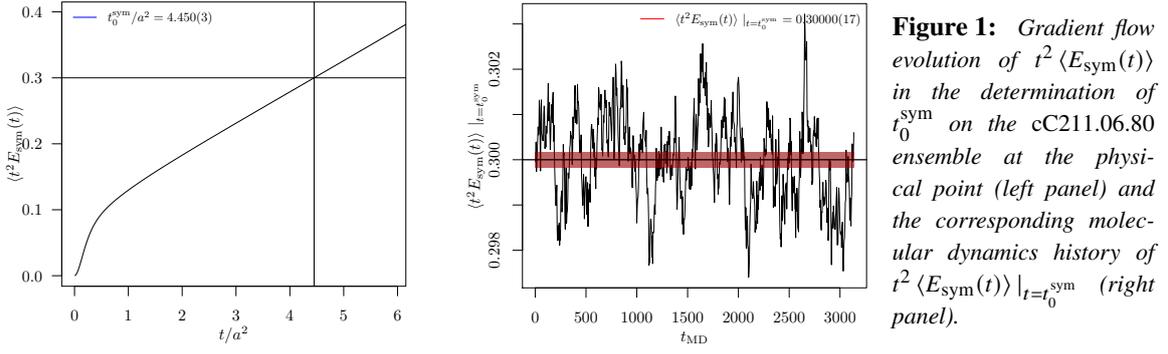

  \centering
  \includegraphics[page=6,width=0.35\textwidth]{{plots/cC211.06.80_combined.gradflow}}\qquad
  \includegraphics[page=7,width=0.35\textwidth]{{plots/cC211.06.80_combined.gradflow}}
  \caption{\it \footnotesize Gradient flow evolution of $t^2 \Esym$ in the determination of $\tsym$ on the \emph{cC211.06.80} ensemble at the physical point (left panel) and the corresponding molecular dynamics history of $t^2 \Esym |_{t = \tsym}$ (right panel).\label{fig:t0sym_evolution}}
\end{SCfigure}

We employ the gradient flow using the Wilson gauge action and use a third-order Runge-Kutta algorithm as proposed in Ref.~\cite{Luscher:2010iy} to evolve the gauge field along the flow time $t/a^2$. 
For the definition of the energy density $\braket{E(t)}$ in our observables, we make use of the clover discretisation of the field tensor, using the notation $\Esym$ in what follows.
In \Cref{fig:t0sym_evolution}, we show the evolution of $t^2 \Esym$ and the corresponding MD history of the observable at the point $t = \tsym$ on ensemble \emph{cC211.06.80} at the physical point as a representative example across our ensemble landscape.

\begin{figure}
  \includegraphics[width=0.3\textwidth,page=1]{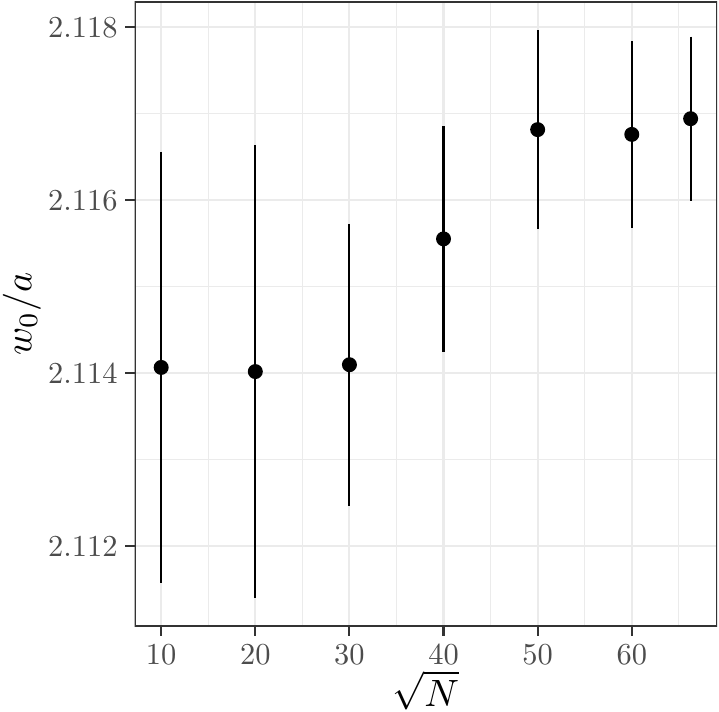}\hfill
  \includegraphics[width=0.3\textwidth,page=2]{plots/gf_error_scaling}\hfill
  \includegraphics[width=0.3\textwidth,page=3]{plots/gf_error_scaling}
  \caption{\it \footnotesize Scaling with the number of trajectories $N$ (of length $\tau=1.5$) of $w_0/a$ (left), its statistical error (middle) and the estimate of the integrated autocorrelation time (right, in units of $\tau=1.0$ trajectories) in an analysis on ensemble \emph{cB211.14.64}. The blue line in the middle panel is a linear fit in the range $30 \leq \sqrt{N} \leq 60$. \label{fig:w0_error_scaling}}
\end{figure}

We use the Gamma method~\cite{Wolff:2003sm} to estimate our statistical errors.
While we do not estimate the exponential tails~\cite{Schaefer:2010hu} of the autocorrelation function of our gradient flow observables, we see good error scaling and stable estimates of the integrated autocorrelation time.
This is shown exemplarily in \Cref{fig:w0_error_scaling} for the \emph{cB211.14.64} ensemble at a pion mass of around $190$~MeV, where we give the evolution as a function of the number of trajectories $N$ (of length $\tau=1.5$) of the observable $w_0/a$, its statistical error and the corresponding estimate of the integrated autocorrelation time.
These appear to be reliable from around $\sqrt{N} \sim 30$ onwards.

The results for all observables using the clover discretisation are given in \Cref{tab:GF_scales}, where we make use of the shorthand notation $s_0 = \sqrt{t_0}$.

\begin{table}
\centering
\begin{adjustbox}{width=\textwidth}
\begin{tabular}{||l|r|r| llll|llll ||}
 \hline
  ensemble & $N_{\text{traj}}$ & $N_{\text{meas}}$ & $\ssym/a$ & $\wsym/a$ & $(\tsym/\wsym)/a$ & $\ssym/\wsym$ & $\tau_\textrm{int}^{s_0}$ &  $\tau_\textrm{int}^{w_0}$ &  $\tau_\textrm{int}^{t_0/w_0}$ &  $\tau_\textrm{int}^{s_0/w_0}$     \\
\hline
  cA211.53.24  & 4488  & 1122 & 1.5306(21) & 1.7597(43) & 1.33139(89)   & 0.86982(100)  & 23(6) & 25(7)   &   7(1)   &  18(4)  \\ 
  cA211.40.24  & 4876  & 1219 & 1.5384(18) & 1.7766(33) & 1.33213(96)   & 0.86592(64)   & 20(5) & 18(4)   &   7(1)   &   9(2)  \\
  cA211.30.32  & 10236 & 2559 & 1.5460( 9) & 1.7928(17) & 1.33314(47)   & 0.86233(32)   & 22(5) & 21(4)   &   9(1)   &  10(2)  \\
  cA211.12.48  & 2608  & 326 & 1.5614(22) & 1.8249(33) & 1.33590(155)  & 0.85559(29)   & 69(30) & 63(27) & 59(25)   &  16(5)  \\
  \hline
  cB211.25.24  & 4580  & 1145 & 1.7937(22) & 2.0992(46) & 1.53260(108)  & 0.85445(77)   & 21(5) & 25(6)   &   5(1)   &   12(2)  \\
  cB211.25.32  & 3960  & 990 & 1.7922(19) & 2.0991(47) & 1.53018(72)   & 0.85380(91)   & 35(10) & 45(14)   &   6(1)   &  28(7)  \\
  cB211.25.48  & 4700  & 1175 & 1.7915( 8) & 2.0982(19) & 1.52966(41)   & 0.85384(38)   & 28(8) & 31(9)   &   9(2)   &  20(5)  \\
  cB211.14.64  & 4952  & 619 & 1.7992( 5) & 2.1175(11) & 1.52875(23)   & 0.84968(23)   & 30(8) & 32(9)   &   8(1)   &  23(6)  \\
  cB211.072.64 & 3065  & 191 & 1.8028( 8) & 2.1272(19) & 1.52784(42)   & 0.84750(41)   & 45(18) & 52(22) &  16(5)   &  41(16) \\
  \hline
  cC211.06.80  & 3140 & 785 & 2.1094( 8) & 2.5045(17) & 1.77670(37)   & 0.84226(27)   & 46(17) & 42(16) &  14(3)   &  26(8)  \\
\hline
\end{tabular}
\end{adjustbox}
  \caption{\label{tab:GF_scales} \it \footnotesize GF scales from the symmetrized action density and corresponding integrated autocorrelation times in units of trajectories of length $\tau=1.0$. The $N_{\text{meas}}$ measurements on each ensemble were performed using different separations in terms of trajectories and the $\tau_{\mathrm{int}}$ were scaled appropriately. Similarly, for the cB211.25.24, cB211.25.32 and cB211.14.64 ensembles, the $\tau_{\mathrm{int}}$ were scaled to take into account the $\tau=1.5$ trajectory lengths used there.}
\end{table}

\section{Extrapolation to the Physical Point}

\begin{figure}
  \includegraphics[width=0.49\textwidth]{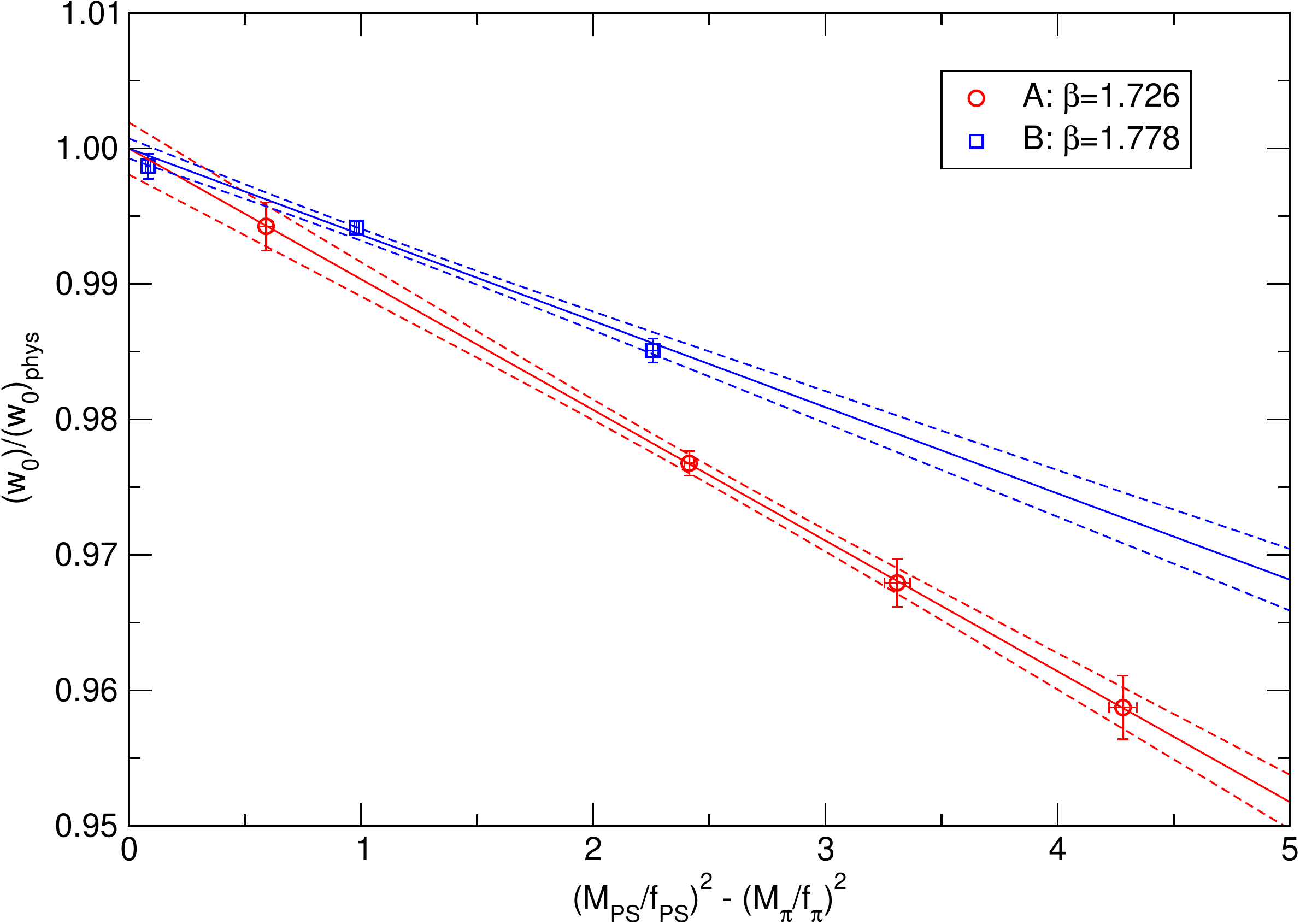}\hfill
\begin{adjustbox}{width=0.49\textwidth,valign=m,margin*=0 2.5cm 0 2cm}
\footnotesize
\centering
\begin{tabular}{||c||c|c|c||}
\hline
$\beta$ & $\wsym / a$ & $\sqrt{\tsym} / a$ & $\tsym / (\wsym a)$\\ \hline
\hline
$~1.726~$ & $~1.8352~(35)~$ & $~1.5660~(22)$ & $~1.3359~(12)~$\\ \hline
$~1.778~$ & $~2.1299~(16)~$ & $~1.80396~(68)$ & $~1.52789~(33)~$\\ \hline
$~1.836~$ & $~2.5045~(17)~$ & $~2.1094~(8)$ & $~1.77670~(37)~$\\ \hline
\end{tabular}
\end{adjustbox}\vspace{-0.8cm}

  \caption{\label{fig:w0_chiral} \it \footnotesize Extrapolation to the physical sea light quark mass of $w_0/a$ at each lattice spacing using \Cref{eq:w0_chiral} (left) and resulting values of all the relative scales at the physical point (right).}
\end{figure}

Before we use the relative scales for further analysis, we follow Ref.~\cite{Bar:2013ora} and extrapolate to the physical light quark mass at each lattice spacing.
Since we have fixed the sea strange and charm quark masses to their physical values to within a few percent, we only parameterise the light quark mass dependence via
\begin{equation}
  w_0 / w_0^{\mathrm{phys}}(\beta) = 1 + c_\beta \cdot \left[ {\left( M_{\mathrm{PS}} / f_{\mathrm{PS}} \right)}^2 - {\left( M_\pi^{\mathrm{iso}} / f^{\mathrm{iso}}_\pi \right)}^2 \right] \label{eq:w0_chiral} \,,
\end{equation}
where $w_0^\mathrm{phys}/a$ and $c_\beta$ are fit parameters and where the pion mass $M_{\mathrm{PS}}$ and pion decay constant $f_{\mathrm{PS}}$ have been corrected for finite size effects as detailed in Ref.~\cite{ExtendedTwistedMass:2021qui}.
The quantities $M_\pi^{\mathrm{iso}}$ and $f_\pi^{\mathrm{iso}}$ correspond to these quantities in the isosymmetric limit of QCD.
The quality of the fit is shown for $\wsym/a$ in the left panel of \Cref{fig:w0_chiral} and the resulting values of all the relative scales at the physical point are given in the right panel.
While we cannot perform this fit for our finest lattice spacing, the single ensemble there is very close to the physical point and we simply use the relative scales as they are.

\section{Setting the Scale}

\begin{SCfigure}
  \includegraphics[width=0.6\textwidth]{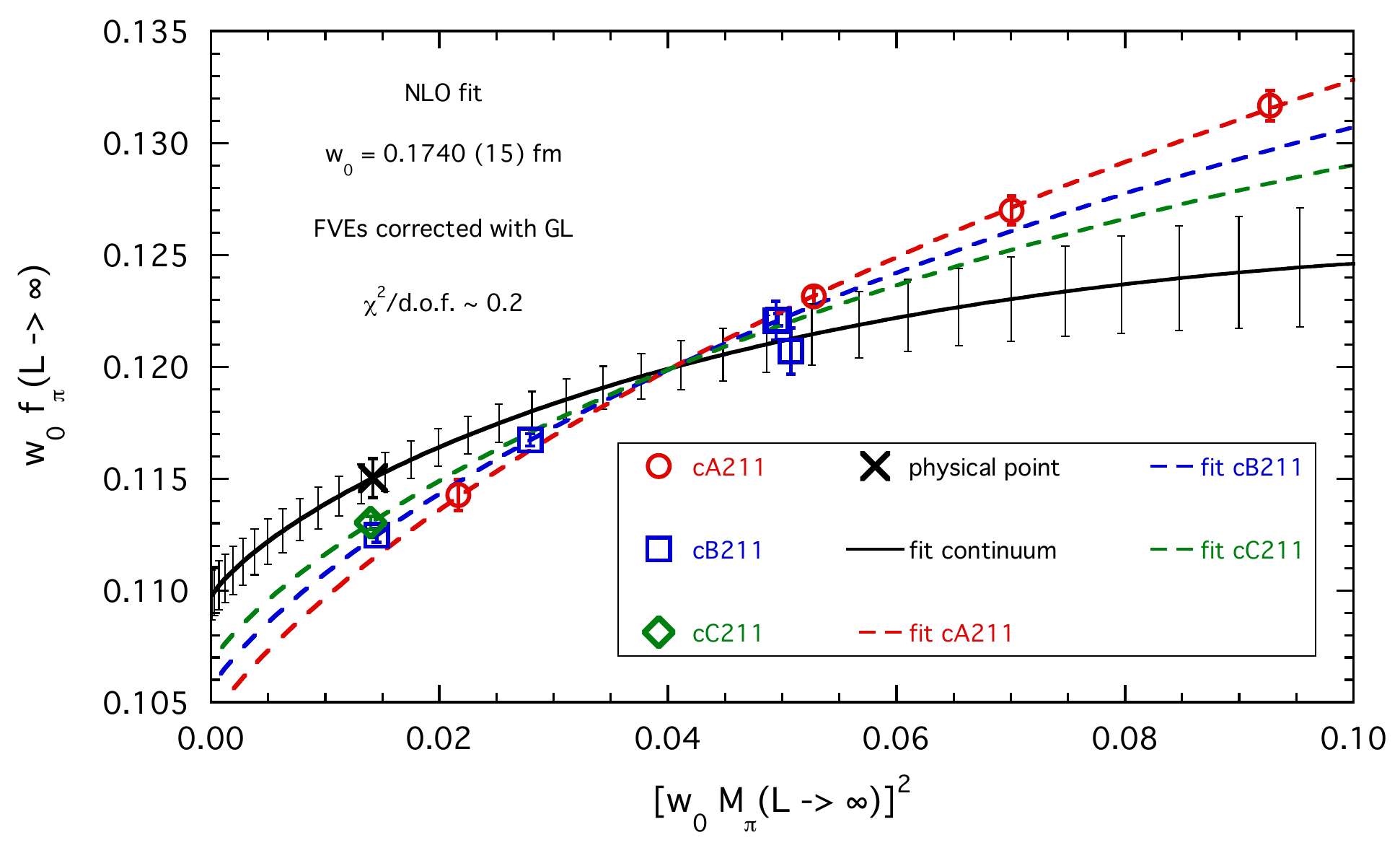}
  \caption{\label{fig:fit_fPi_NLO} \it \footnotesize Fit of \Cref{eq:fpi_NNLO} with $A_2=0$ to the data for $w_0 f_\pi$.}
\end{SCfigure}

We first attempt to set the scale via the pion decay constant directly, fitting the data for $w_0 f_\pi(L \to \infty)$ (which has been corrected for finite size effects) using the following functional form
\begin{equation}
     w_0 f_\pi(L \to \infty) = w_0 f \left[ 1 - 2 \xi \mbox{log}(\xi) + 2 A_1 \xi + A_2 \xi^2 + \frac{a^2}{w_0^2} \left( D_0 + D_1 \xi \right) \right] \,,
     \label{eq:fpi_NNLO}
\end{equation}
where $\xi$ is defined as
\begin{equation}
   \xi \equiv \frac{M_\pi^2(L \to \infty)}{(4\pi f)^2} = \frac{(w_0 M_\pi)^2}{(4\pi w_0 f)^2} ~ 
                      \frac{1}{\left[ 1 - \frac{1}{4} \Delta_{\mathrm{FVE}}^\pi(L) \right]^2} \,, 
   \label{eq:xi_ell_meson}
\end{equation}
and where, with respect to a pure NLO ansatz, we have added a possible higher-order term quadratic in $\xi$ as well as discretization effects proportional to $a^2$ and $a^2 M_\pi^2$.
For details we refer to Ref.~\cite{ExtendedTwistedMass:2021qui} and show the fit with $A_2=0$ in \Cref{fig:fit_fPi_NLO} with the result $w_0 = 0.1740(15)$ fm, corresponding to an $0.8$\% error.

\begin{SCfigure}
  \centering
  \includegraphics[width=0.8\textwidth]{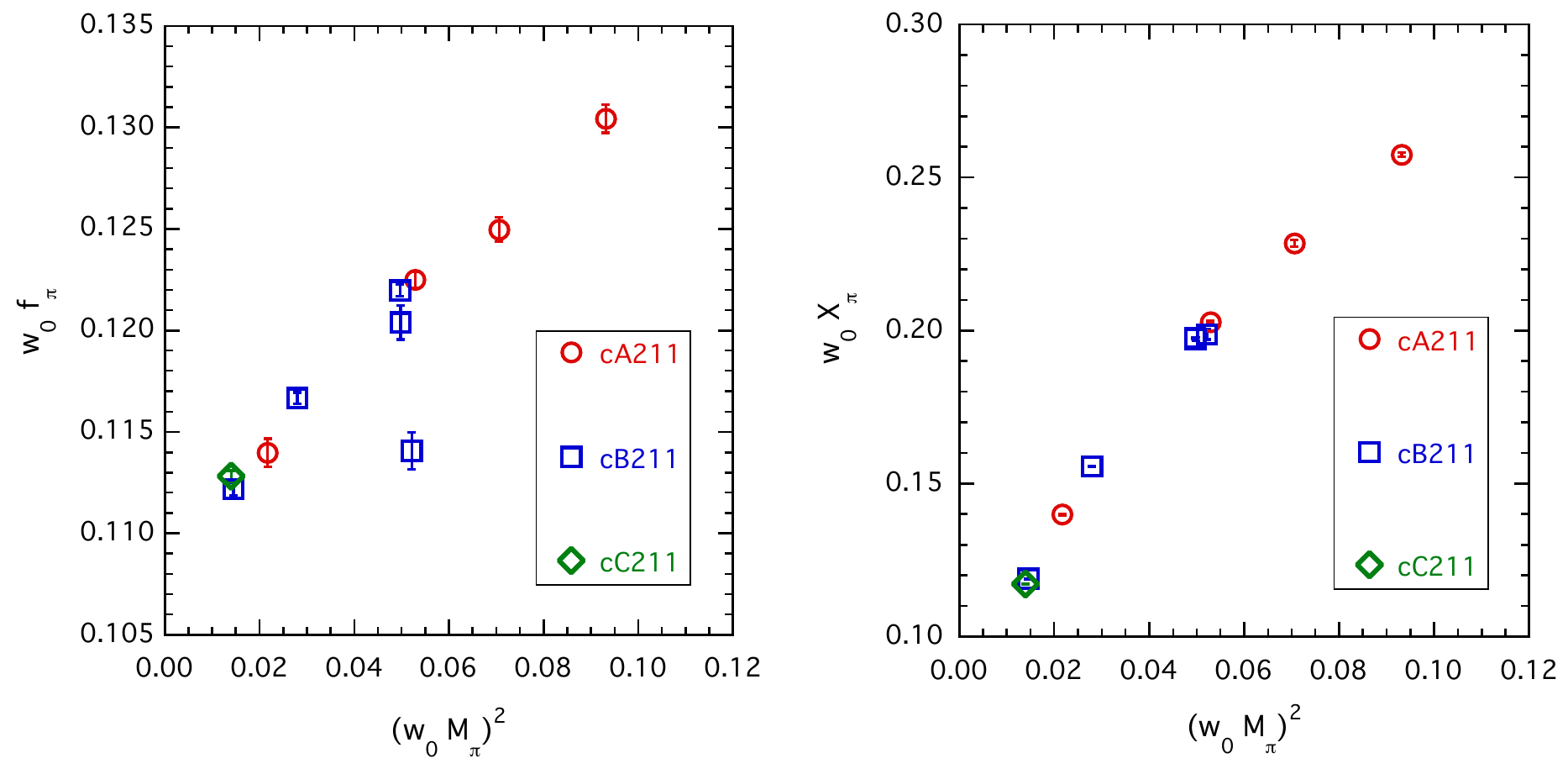}
  \caption{\label{fig:pion_data} \it \footnotesize Raw data for the quantities $w_0 f_\pi$ (left) and $w_0 X_\pi$ (right, defined in \Cref{eq:Xpi}).}
\end{SCfigure}

\begin{figure}
  \hspace{-0.07\textwidth}
  \includegraphics[height=5.5cm]{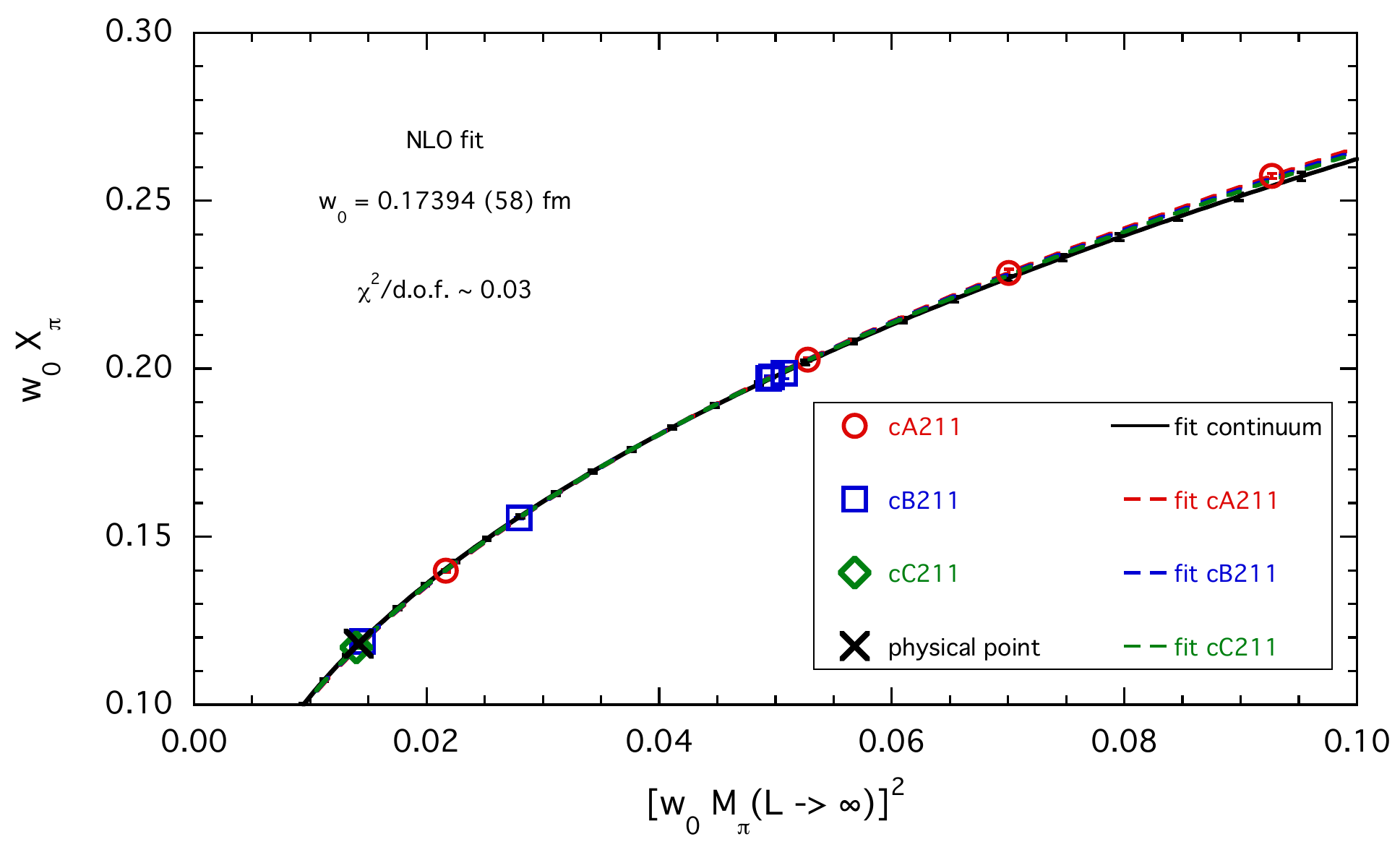}
  \includegraphics[height=5.5cm]{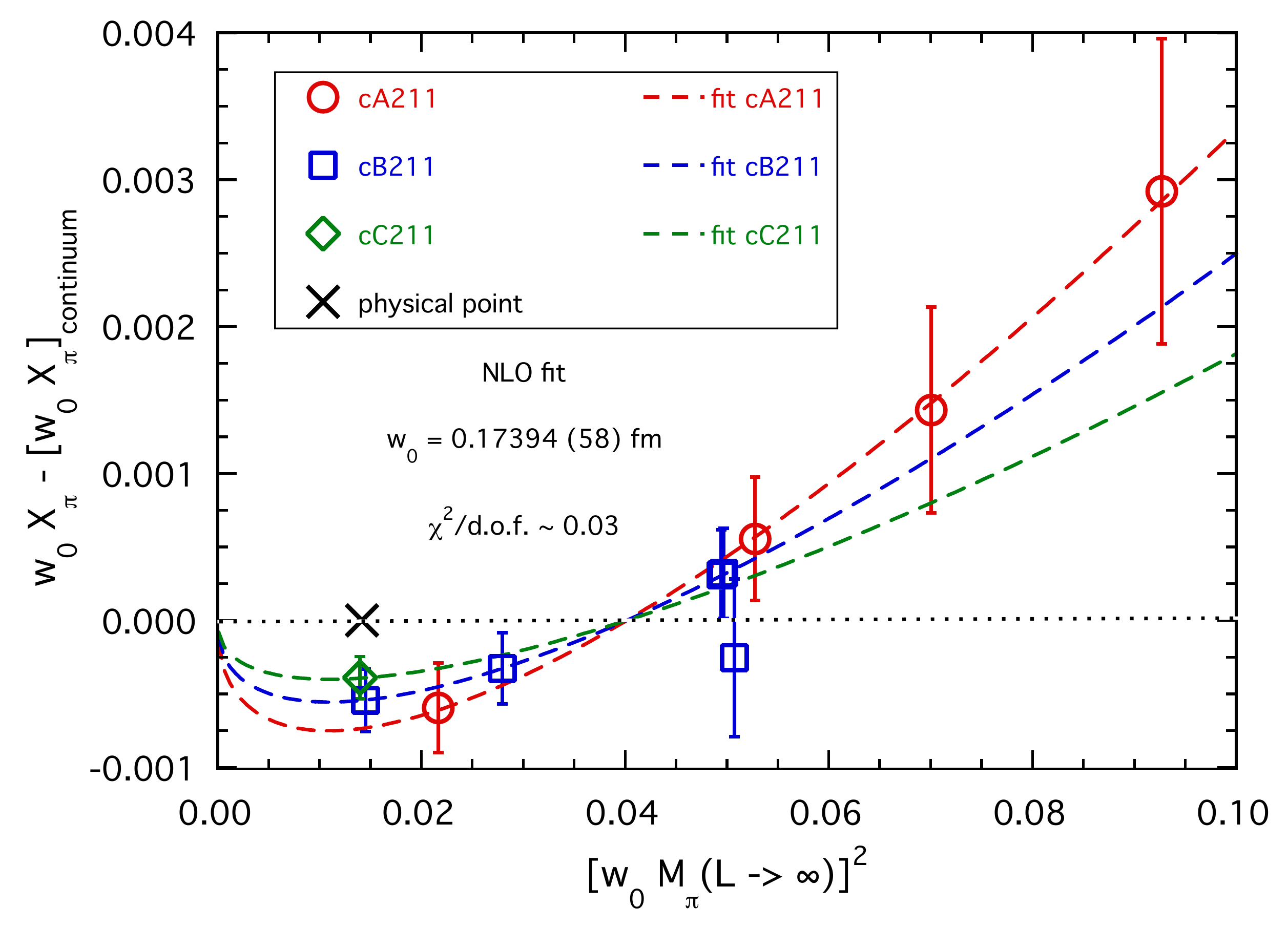}
  \caption{\label{fig:XPi_NLO} \it \footnotesize Fit of \Cref{eq:XPi_fit} to the data for $w_0 X_\pi = w_0 {\left( f_\pi M_\pi^4 \right)}^{(1/5)}$ (left) and detail with the resulting continuum curve subtracted to better visualise the very small residual lattice artefacts (right).}
\end{figure}

In order to better exploit statistical correlations in the data for $f_{\mathrm{PS}}$ and $M_{\mathrm{PS}}$ as well as cancellations of discretisation and finite size effects, we consider the quantity 
\begin{equation}
  \label{eq:Xpi}
  X_\pi = \left( f_\pi M_\pi^4 \right)^{1/5} \,,
\end{equation}
for which we compare the raw data for $w_0 f_{\mathrm{PS}}$ in the left panel of \Cref{fig:pion_data} to the raw data for $w_0 X_{\mathrm{PS}}$ in the right panel.
It is clear that especially the finite size effects are greatly reduced in this combination. 
We proceed to fit
\begin{equation}
\begin{split}
    \label{eq:XPi_fit}
    w_0 X_\pi = (w_0 f)  \left\{ (4 \pi)^4 \xi^2 \left[ 1 - 2 \xi \mbox{log}(\xi) + 2 A_1 \xi + A_2^\prime \xi^2  + a^2 \left( D_0^\prime + D_1^\prime \xi \right) \right] \right\}^{1/5} \\
                     \cdot \left( 1 + F_{\mathrm{FVE}} ~ \xi^2 e^{-M_\pi L} / (M_\pi L)^{3/2} \right) \,,
\end{split}
\end{equation}
the result of which is shown in in the left panel of \Cref{fig:XPi_NLO}.
In the right panel, instead, we have subtracted the resulting continuum curve to better visualise the residual lattice artefacts which are very small and yet very well captured by the fit.

For details we again refer to Ref.~\cite{ExtendedTwistedMass:2021qui}, where different cuts in the data and variations of the higher order terms are used to obtain estimates of systematic errors.
Repeating the fits for the scales $w_0$, $t_0$ and $t_0/w_0$, we obtain
\begin{align}
  w_0 & = 0.17383 ~ (57)_{\rm stat+fit} ~ (26)_{\rm syst} ~ [63] ~ \mbox{fm} \label{eq:w0} \,,\\
   \sqrt{t_0} & = 0.14436 ~ (54)_{\rm stat+fit} ~ (30)_{\rm syst} ~ [61] ~ \mbox{fm} \label{eq:s0} \,,\\
    t_0 / w_0 & = 0.11969 ~ (52)_{\rm stat+fit} ~ (33)_{\rm syst} ~ [62] ~ \mbox{fm} \label{eq:t0_ov_w0} \,,
\end{align}
with errors added in quadrature and given in square brackets, resulting in an improvement in precision by a factor of about $2.5$ compared to the determination from $f_\pi$.

\begin{SCfigure}[0.65]
  \centering
  \includegraphics[width=0.56\textwidth]{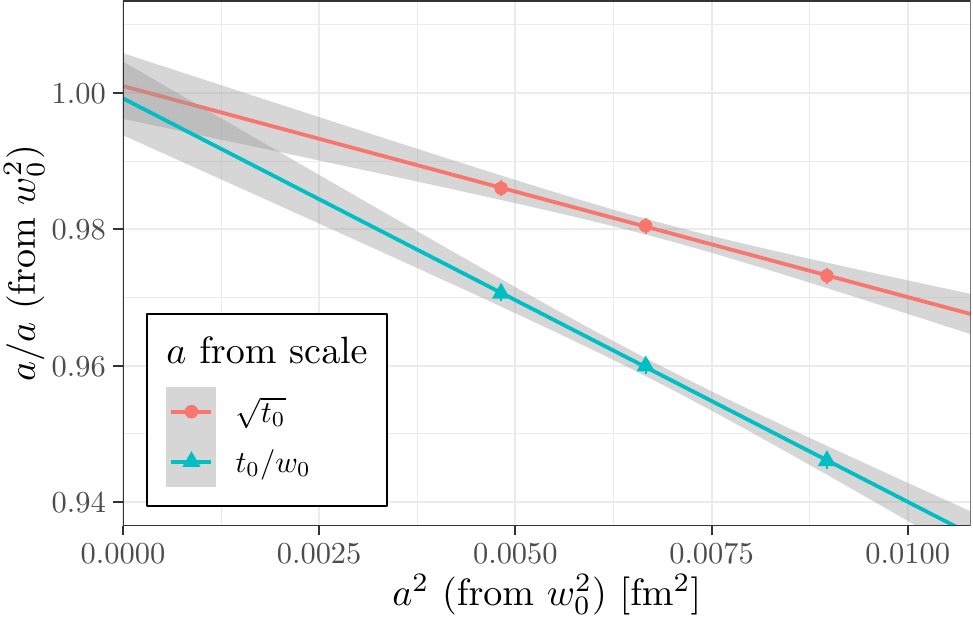}
  \hspace{0.005\textwidth}
\caption{\label{fig:rel_lat_spac} \it \footnotesize Ratios of lattice spacings determined via $\sqrt{t_0}$ and $t_0/w_0$ with the lattice spacing determined via $w_0$ with linear fits superimposed.}
\end{SCfigure}

The values of the lattice spacing $a$ corresponding to \Cref{eq:w0,eq:s0,eq:t0_ov_w0} are given in \Cref{tab:spacings}.
These three determinations of $a$ differ by ${\cal{O}}(a^2)$ effects, which can be parameterised in their ratios by a function linear in $a^2$, as shown in Fig.~\ref{fig:rel_lat_spac}.
In particular, we get: $a(\sqrt{t_0}) / a(w_0) \simeq 1 -  0.09\,(2) ~ a^2(w_0) / w_0^2$ and $a(t_0 / w_0) / a(w_0) \simeq 1 - 0.18\,(2) ~ a^2(w_0) / w_0^2$, consistent with $a^2$-scaling.

\begin{SCtable}
{\footnotesize
\begin{tabular}{|c||c|c|c|}
\hline
scale & $a_A(\beta=1.726)$ & $a_B(\beta=1.778)$ & $a_C(\beta=1.836)$ \\ \hline
\hline
$w_0$         & 0.09471(39) & 0.08161(30) & 0.06941(26)\\ \hline
$\sqrt{t_0}$  & 0.09217(41) & 0.08002(34) & 0.06844(29)\\ \hline
$t_0 / w_0$   & 0.08960(47) & 0.07834(41) & 0.06737(35)\\ \hline
\end{tabular}
}
  \caption{\it \footnotesize Values of the lattice spacing $a$ (in fm) corresponding to the three GF scales $w_0$, $\sqrt{t_0}$, $t_0 / w_0$ and to the corresponding relative scales given in the right panel of \Cref{fig:w0_chiral}.}
\label{tab:spacings}
\end{SCtable}

\section{The ratio $f_K/f_\pi$}

\begin{SCfigure}
  \includegraphics[width=0.6\textwidth]{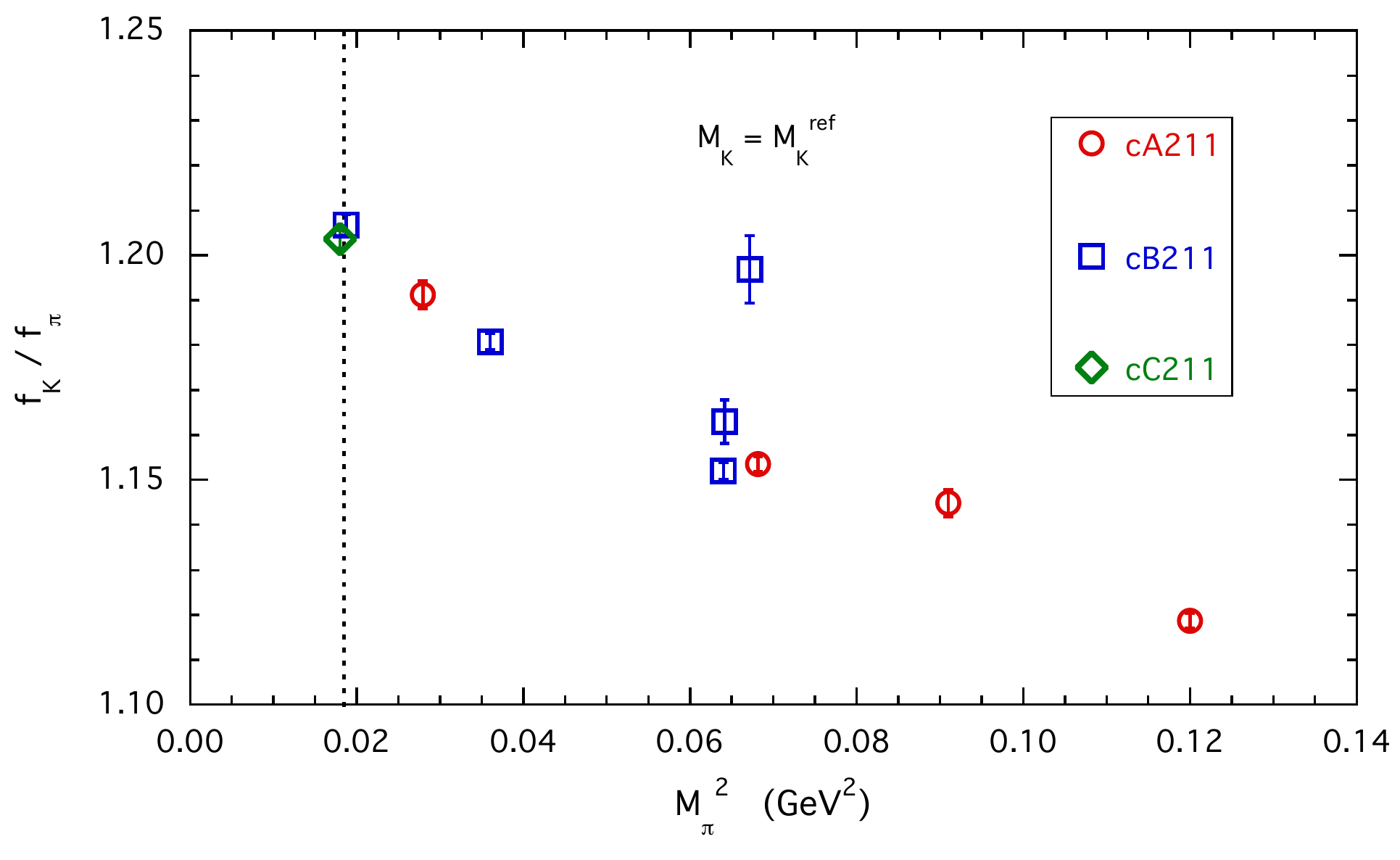}
  \caption{\label{fig:fKPi_data} \it \footnotesize Data for $f_K/f_\pi$ on all ensembles used in this work interpolated to the reference kaon mass $ {(M_K^\mathrm{ref})}^2 = (M_K^{\mathrm{iso}})^2 + {( M_\pi^2 - M_\pi^{\mathrm{iso}} )}^2/2$.}
\end{SCfigure}

Finally, we employ the lattice spacing determined via $w_0/a$ and fixed by $X_\pi$ to interpolate our data for $f_K/f_\pi$ to a reference kaon mass ${(M_K^\mathrm{ref})}^2 = (M_K^{\mathrm{iso}})^2 + {( M_\pi^2 - M_\pi^{\mathrm{iso}} )}^2/2$.
This interpolated data for $f_K/f_\pi$ is shown in \Cref{fig:fKPi_data}.
We further apply finite size corrections as detailed in Ref.~\cite{ExtendedTwistedMass:2021qui} and the data using the Ansatz
\begin{equation}
     \frac{f_K}{f_\pi}(L \to \infty) = R_0 \left[ 1 + \frac{5}{4} \xi \mbox{log}(\xi) + R_1 \xi + R_2 \xi^2 + \frac{a^2}{w_0^2} \left( \widetilde{D}_0 + \widetilde{D}_1 \xi \right) \right] \,.
     \label{eq:fKPi_fit}  
\end{equation}

\begin{SCfigure}
  \includegraphics[width=0.6\textwidth]{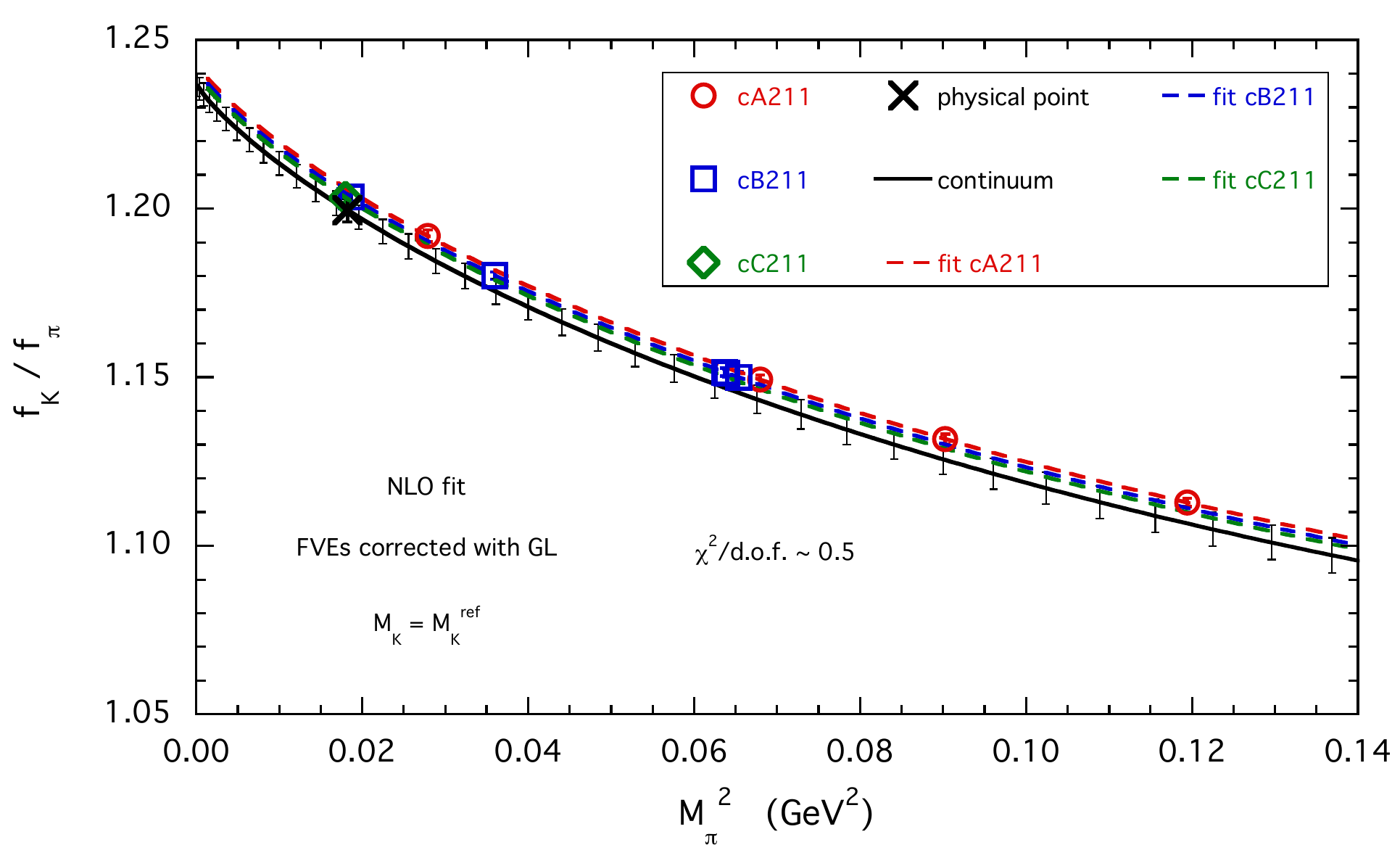}
  \caption{\label{fig:fKPi_NLO} \it \footnotesize Fit of \Cref{eq:fKPi_fit} to the $f_K/f_\pi$ data shown in \Cref{fig:fKPi_data} corrected for finite size effects.}
\end{SCfigure}

As shown in \Cref{fig:fKPi_NLO}, this results in an excellent fit and we obtain
\begin{equation}
  \label{eq:fKPi_isoQCD}
  \left( \frac{f_K}{f_\pi} \right)^{\mathrm{iso}} = 1.1995 ~ (44)_{\textrm{stat+fit}} ~ (7)_{\textrm{syst}} ~ [44] \,, 
\end{equation}
at the physical point in the isosymmetric limit of QCD, where again, estimates of the systematic errors are obtained by performing different types of fits and data cuts. 

\section{Conclusions and Outlook}

\begin{figure}
  \includegraphics[height=6.5cm]{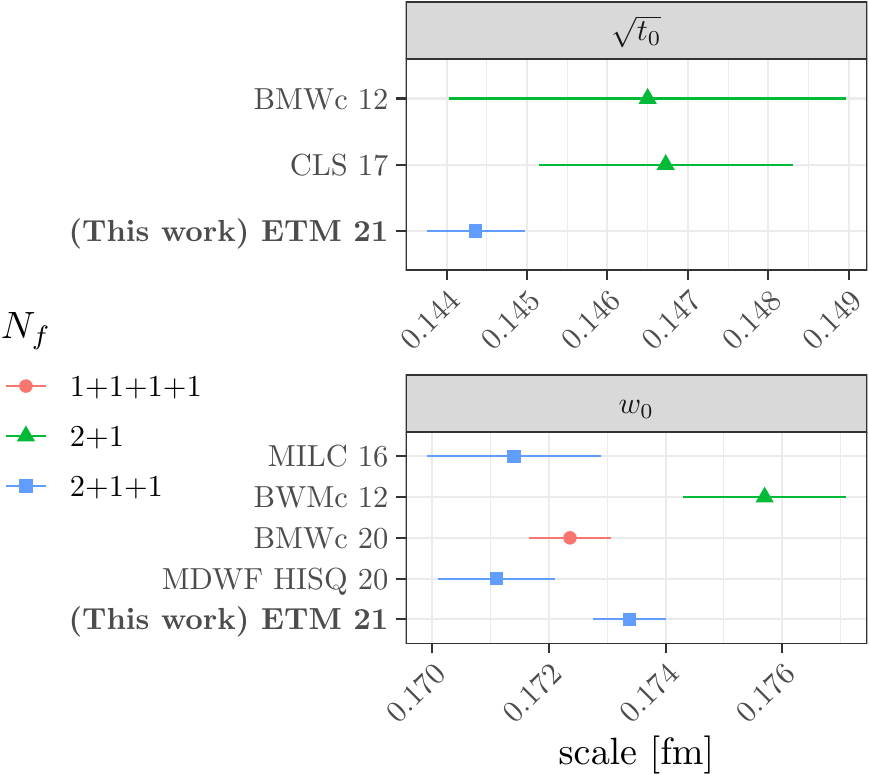}
  \includegraphics[height=6.5cm]{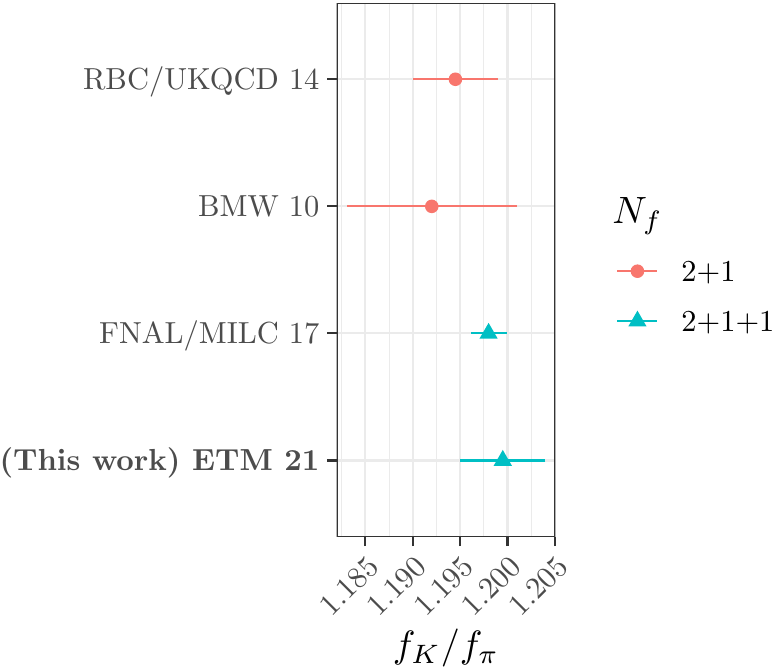}
  \caption{\label{fig:comparison} \it \footnotesize Comparison of the results of this work for the scales $w_0$ and $\sqrt{t_0}$ from Refs.~\cite{Borsanyi:2012zs,Bruno:2016plf,Miller:2020evg,Borsanyi:2020mff,MILC:2015tqx} (left panel) and for $f_K/f_\pi$ to those from Refs.~\cite{PhysRevD.81.054507,PhysRevD.93.074505,Bazavov:2017lyh} (right panel).}
\end{figure}

We conclude by comparing our results for the scales $w_0$ and $\sqrt{t_0}$ to an incomplete selection from Refs.~\cite{Borsanyi:2012zs,Bruno:2016plf,Miller:2020evg,Borsanyi:2020mff,MILC:2015tqx} in the left panel of \Cref{fig:comparison} and our result for $f_K/f_\pi$ to a selection from Refs.~\cite{PhysRevD.81.054507,PhysRevD.93.074505,Bazavov:2017lyh} in the right panel.
For more complete comparisons we refer to the FLAG review~\cite{Aoki:2021kgd}.
In future publications we plan to extend the set of ensembles by simulations at a fourth lattice spacing, several more volumes and further values of the light sea quark mass at $\beta=1.836$.
An alternative scale setting employing the mass of the omega baryon is currently being studied with the aim of also including QED effects.

\acknowledgments

{\footnotesize
We thank all the ETMC members for a very productive collaboration.

We acknowledge PRACE for access to Marconi and Marconi100 at CINECA under the grants Pra17-4394, Pra20-5171 and Pra22-5171, and CINECA for providing us CPU time under the specific initiative INFN-LQCD123. We also acknowledge PRACE for awarding us access to HAWK, hosted by HLRS, Germany, under the grant with id 33037.
The authors gratefully acknowledge the Gauss Centre for Supercomputing e.V.~(www.gauss-centre.eu) for funding the project pr74yo by providing computing time on the GCS Supercomputer SuperMUC at Leibniz Supercomputing Centre (www.lrz.de).
Some of the ensembles for this study were generated on Jureca Booster~\cite{jureca} and Juwels~\cite{JUWELS} at the J{\"u}lich Supercomputing Centre (JSC) and we gratefully acknowledge the computing time granted there by the John von Neumann Institute for Computing (NIC).
We also acknowledge access to the Bonna HPC Cluster at the University of Bonn.

The project has received funding from the Horizon 2020 research and innovation program of the European Commission under the Marie Sklodowska-Curie grant agreement No 642069 (HPC-LEAP) and under grant agreement No 765048 (STIMULATE).
The project was funded in part by the NSFC (National Natural Science Foundation of China) and the DFG (Deutsche Forschungsgemeinschaft, German Research Foundation) through the Sino-German Collaborative Research Center grant TRR110 ``Symmetries and the Emergence of Structure in QCD'' (NSFC Grant No. 12070131001, DFG Project-ID 196253076 - TRR 110).

R.F. acknowledges support from the University of Tor Vergata through the Grant “Strong Interactions: from Lattice QCD to Strings, Branes and Holography” within the Excellence Scheme “Beyond the Borders”
F.S.~and S.S.~are supported by the Italian Ministry of Research (MIUR) under grant PRIN 20172LNEEZ.
F.S.~is supported by INFN under GRANT73/CALAT.
P.D.~acknowledges support form the European Unions Horizon 2020 research and innovation programme under the Marie Sk\l{}odowska-Curie grant agreement No. 813942 (EuroPLEx) and from INFN.~under the research project INFN-QCDLAT.
S.B.~and J.F.~are supported by the H2020 project PRACE6-IP (grant agreement No 82376) and the COMPLEMENTARY/0916/0015 project funded by the Cyprus Research Promotion Foundation.
The authors acknowledge support from project NextQCD, co-funded by the European Regional Development Fund and the Republic of Cyprus through the Research and Innovation Foundation (EXCELLENCE/0918/0129).
}

\bibliographystyle{JHEP}
{\footnotesize

\providecommand{\href}[2]{#2}\begingroup\raggedright\endgroup
 
}

\end{document}